\documentclass[twocolumn,showpacs,preprintnumbers,
amsmath,amssymb,aps,prd,nofootinbib,superscriptaddress,
eqsecnum]{revtex4}
\usepackage{graphicx}

\makeatletter
\renewcommand{\p@subsection}{}
\makeatother

\renewcommand{\thesection}{\arabic{section}}

\renewcommand{\theequation}{\arabic{section}.\arabic{equation}}

\begin{document}

\title{Vector-axialvector mixing
from a chiral effective field theory at finite temperature}

\author{Masayasu Harada}
\affiliation{%
Department of Physics, Nagoya University,
Nagoya, 464-8602, Japan}
\author{Chihiro Sasaki}
\affiliation{%
Physik-Department,
Technische Universit\"{a}t M\"{u}nchen,
D-85747 Garching, Germany
}
\author{Wolfram Weise}
\affiliation{%
Physik-Department,
Technische Universit\"{a}t M\"{u}nchen,
D-85747 Garching, Germany
}

\date{\today}

\begin{abstract}
We study the vector-axialvector mixing in a hot medium
and its evolution toward the chiral phase transition
using different symmetry restoration scenarios based on
the generalized hidden local symmetry framework.
We show that the presence of the $a_1$ meson 
reduces the vector spectral function around $\rho$ meson
mass and enhances it around $a_1$ meson mass.
The coupling strength of $a_1$ to $\rho$ and $\pi$ vanishes 
at the critical temperature due to the degenerate $\rho$-$a_1$ 
masses. This feature holds rigorously in the chiral limit
and still stays intact to good approximation
for the physical pion mass.
\end{abstract}


\maketitle

\section{Introduction}

In-medium changes of hadron properties are considered to be 
indicators of the tendency towards chiral symmetry restoration 
in hot and/or dense QCD. 
In particular, the short-lived vector mesons like the $\rho$ 
mesons are expected to carry information on the modifications 
of hadrons in matter~\cite{review}. In the presence of hot matter 
the vector and axialvector current correlators are mixed due 
to pions in the heat bath. At low temperatures this process is 
described in a model-independent way in terms of a low-energy 
theorem based on chiral symmetry~\cite{theorem}. The vector spectral
function is then modified by axialvector mesons through the 
mixing theorem~\cite{vamix}.

The validity of this theorem is, however, limited to temperatures 
$T \ll 2f_\pi$, where $f_\pi$ is the pion decay constant in vacuum.
At higher temperatures hadrons other than pions are thermally 
activated. Thus one needs in-medium correlators 
systematically involving those excitations.

In this paper
we show the effects of the mixing (hereafter V-A mixing), and how the 
axialvector mesons affect the spectral function near the chiral 
phase transition, within an effective field theory.
Our analysis will be carried out assuming several possible patterns 
of chiral symmetry restoration: dropping or non-dropping $\rho$
meson mass along with changing $a_1$ meson mass, 
both considered to be options from a
phenomenological point of view. The effect of explicit
chiral symmetry breaking is also examined.

\section{Generalized hidden local symmetry}

Several models exist which explicitly include the axialvector 
meson in addition to the pion and vector meson consistently with 
the chiral symmetry of QCD, such as the Massive Yang-Mills 
model~\cite{MassiveYM}, the anti-symmetric tensor field 
method~\cite{Gasser:1983yg} and the approach based on 
generalized 
hidden local symmetry (GHLS)~\cite{ghls:pr,Kaiser:1990yf}.
These models are equivalent~\cite{Kaiser:1990yf,equi,hy:pr} for 
tree-level amplitudes in the low-energy limit.

\subsection{Lagrangian} 

The GHLS Lagrangian is based on
a $G_{\rm{global}} \times G_{\rm{local}}$ symmetry,
where $G_{\rm global}=[SU(N_f)_L \times SU(N_f)_R]_{\rm global}$ 
is the chiral symmetry and 
$G_{\rm local}=[SU(N_f)_L \times SU(N_f)_R]_{\rm local}$ 
is the GHLS.
The whole symmetry $G_{\rm global}\times G_{\rm local}$
is spontaneously broken to a diagonal $SU(N_f)_V$.
The basic quantities are
the GHLS gauge bosons, $L_\mu$ and $R_\mu$,
identified with the vector and axialvector mesons as
$V_\mu = (R_\mu + L_\mu)/2$ and 
$A_\mu = (R_\mu - L_\mu)/2$,
and 
three matrix valued variables $\xi_L$, $\xi_R$
and $\xi_M$, which are combined in a
$N_f \times N_f$ special-unitary matrix
$U = \xi_L^\dagger \xi_M \xi_R$.

The fundamental objects are the Maurer-Cartan 1-forms
defined by
\begin{eqnarray}
\hat{\alpha}_{L,R}^\mu = D^\mu\xi_{L,R}\cdot\xi_{L,R}^\dagger /i\,,
\quad
\hat{\alpha}_M^\mu = D^\mu\xi_M\cdot\xi_M^\dagger /(2i)\,,
\end{eqnarray}
where
the covariant derivatives of $\xi_{L,R,M}$ are given by
\begin{eqnarray}
&&
D_\mu \xi_L 
 = \partial_\mu\xi_L - iL_\mu\xi_L + i\xi_L{\cal{L}}_\mu\,,
\nonumber\\
&&
D_\mu \xi_R 
 = \partial_\mu\xi_R - iR_\mu\xi_R + i\xi_R{\cal{R}}_\mu\,,
\nonumber\\
&&
D_\mu \xi_M 
 = \partial_\mu\xi_M - iL_\mu\xi_M + i\xi_M R_\mu\,,
\end{eqnarray}
with ${\cal{L}}_\mu$ and ${\cal{R}}_\mu$ being the external
gauge fields introduced by gauging $G_{\rm{global}}$.
There are four independent terms with lowest derivatives:
\begin{eqnarray}
&&
{\cal L}_V 
 = F^2 \mbox{tr}\bigl[ \hat{\alpha}_{\parallel\mu}
   \hat{\alpha}_\parallel^\mu \bigr]\,,
\quad
{\cal L}_A 
 = F^2 \mbox{tr}\bigl[ \hat{\alpha}_{\perp\mu}
   \hat{\alpha}_\perp^\mu \bigr]\,,
\nonumber\\
&&
{\cal L}_M 
 = F^2 \mbox{tr}\bigl[ \hat{\alpha}_{M\mu}
   \hat{\alpha}_M^\mu \bigr]\,,
\nonumber\\
&&
{\cal L}_\pi 
 = F^2 \mbox{tr}\bigl[ \bigl( \hat{\alpha}_{\perp\mu}
   {}+ \hat{\alpha}_{M\mu} \bigr)
   \bigl( \hat{\alpha}_{\perp}^\mu
   {}+ \hat{\alpha}_{M}^\mu \bigr)\bigr]\,,
\label{lag a-d}
\end{eqnarray}
where $F$ is a parameter of dimension $1$
and 
$\hat{\alpha}_{\parallel,\perp}^\mu
 = \bigl( \xi_M\hat{\alpha}_R^\mu\xi_M^\dagger 
  {}\pm \hat{\alpha}_L^\mu \bigr)/2\,$.
The kinetic term of the gauge bosons 
is given by 
\begin{equation}
{\cal L}_{\rm kin}(L_\mu,R_\mu)
 = {}- \frac{1}{4g^2}\mbox{tr}\bigl[ L_{\mu\nu}L^{\mu\nu}
   {}+ R_{\mu\nu}R^{\mu\nu} \bigr]\,,
\label{lag kin}
\end{equation}
where $g$ is the GHLS gauge coupling 
and the field strengths are defined by
$L_{\mu\nu}
 = \partial_\mu L_\nu - \partial_\nu L_\mu
  {}- i\bigl[ L_\mu, L_\nu \bigr]\,$ and
$R_{\mu\nu}
 = \partial_\mu R_\nu - \partial_\nu R_\mu
  {}- i\bigl[ R_\mu, R_\nu \bigr]\,$.

Combining the terms (\ref{lag a-d}) and (\ref{lag kin}),
the GHLS Lagrangian is given by
\begin{equation}
{\cal L} = a{\cal L}_V + b{\cal L}_A + c{\cal L}_M
 {}+ d{\cal L}_\pi 
 {}+ {\cal L}_{\rm kin}(L_\mu,R_\mu)\,,
\label{lag p^2}
\end{equation}
where $a$, $b$, $c$ and $d$ are dimensionless parameters.
Fields for three types of Nambu-Goldstone (NG) bosons,
$\phi_\sigma, \phi_\perp$ and $\phi_p$, are introduced as
\begin{equation}
\xi_{L,R} = e^{i(\phi_\sigma \mp \phi_\perp)}\,,
\quad
\xi_M = e^{2i\phi_p}\,.
\end{equation}
The pion field $\phi_\pi$ is given by the combination
\begin{equation}
\phi_\pi = \phi_\perp + \phi_p\,,
\end{equation}
while two remaining would-be NG bosons~\cite{ghls},
$\phi_\sigma$ and
\begin{equation}
\phi_q = \frac{1}{b+c}\left( c\phi_p - b\phi_\perp \right)\,,
\end{equation}
representing the longitudinal vector and axialvector
degrees of freedom,
are absorbed into the $\rho$ and $a_1$.
The $\pi, \sigma$ and $q$ fields are normalized by corresponding 
decay constants:
\begin{equation}
\phi_\pi = \frac{\pi}{F_\pi}\,,
\quad
\phi_\sigma = \frac{\sigma}{F_\sigma}\,,
\quad
\phi_q = \frac{q}{F_q}\,.
\end{equation}
The pion decay constant, the meson bare masses and the coupling
strength of the $\rho$ and $a_1$ to the vector and axialvector
currents, $J^\mu$ and $J_5^\mu$, are given by
\begin{eqnarray}
&&
F_\pi^2 = \left( d + \frac{bc}{b+c} \right) F^2\,,
\nonumber\\
&&
M_\rho^2 = g^2 F_\sigma^2 = ag^2F^2\,,
\nonumber\\
&&
M_{a_1}^2 = g^2 F_q^2 = (b+c)g^2F^2\,,
\nonumber\\
&&
g_\rho = agF^2\,,
\quad
g_{a_1} = bgF^2\,.
\label{tree}
\end{eqnarray}

\subsection{Weinberg sum rules} 

The axialvector and vector current correlators are defined as
\begin{eqnarray}
&&
\int d^4x\,e^{iqx}
\left\langle 0 \vert \, T\,
  J_{5}^\mu(x) J_{5}^\nu(0)
\vert 0 \right\rangle
\nonumber\\
&&
\qquad
= G_A(Q^2)(q^\mu q^\nu - q^2 g^{\mu\nu})\,,
\nonumber\\
&&
\int d^4x\,e^{iqx}
\left\langle 0 \vert \, T\,
  J^\mu(x) J^\nu(0)
\vert 0 \right\rangle
\nonumber\\
&&
\qquad
= G_V(Q^2)(q^\mu q^\nu - q^2g^{\mu\nu})\,,
\label{CCdef}
\end{eqnarray}
where $Q^2 = - q^2 > 0$ is the space-like squared momentum.
When these correlators are saturated by the lowest lying mesons
at tree level, we have
\begin{eqnarray}
G_A(Q^2)
= \frac{F_\pi^2}{Q^2} + \frac{F_{a_1}^2}{M_{a_1}^2+Q^2}\,,
\quad
G_V(Q^2)
= \frac{F_\rho^2}{M_\rho^2 + Q^2}\,,
\nonumber\\
\label{pole}
\end{eqnarray}
where the $a_1$ and $\rho$ decay constants are defined by
\begin{eqnarray}
F_{a_1}^2 = \Bigl( \frac{g_{a_1}}{M_{a_1}} \Bigr)^2 
= \frac{b^2}{b+c}F^2\,,
\quad
F_\rho^2 = \Bigl( \frac{g_\rho}{M_\rho} \Bigr)^2 
= a F^2\,.
\nonumber\\
\end{eqnarray}
The same correlators can be evaluated by the operator product
expansion (OPE), which shows that
the difference between two correlators scales as $1/Q^6$~\cite{ope}
~\footnote{
 We assume factorization of four-quark condensates.
}:
\begin{equation}
G_A^{\rm(OPE)}(Q^2) - G_V^{\rm(OPE)}(Q^2) 
= \frac{32\pi}{9} \frac{\alpha_s \,\langle \bar{q}q\rangle^2}{Q^6}\,.
\label{A-V:OPE}
\end{equation}

We require that the high energy behavior of the difference between
the two correlators in the GHLS agrees with that in the OPE:
$G_A(Q^2) - G_V(Q^2)$ approaches $\sim 1/Q^6$.
This condition is satisfied only if the following relations hold:
\begin{eqnarray}
F_\pi^2 + F_{a_1}^2 = F_\rho^2\,,
\quad
F_{a_1}^2 M_{a_1}^2 = F_\rho^2 M_\rho^2\,,
\label{WSR}
\end{eqnarray}
which are nothing but the pole saturated forms of the 
Weinberg first and second sum rules~\cite{Weinberg}.
In terms of the parameters of the GHLS Lagrangian,
the above relations can be traced back to
\begin{equation}
a = b \,, \quad d = 0 \,.
\label{tsl}
\end{equation}

In Ref.~\cite{ghls} it was shown that
the parameter relations are stable
against the renormalization group evolution:
This implies {\it the non-renormalization of the Weinberg sum 
rules expressed in terms of the leading order parameters
in the GHLS}~\footnote
{
 The GHLS Lagrangian does not include scalar $\bar{q}q$ modes
 which are assumed to be heavier than other mesons incorporated.
 This may not be true near the critical point within the 
 Ginzburg-Landau picture of the phase transition. The scalar
 mesons thus modify the renormalization group structure.
}.
In the following studies, we adopt the GHLS Lagrangian with
$a=b$ and $d=0$ as a reliable basis which describes the spectral
function sum rules.

\subsection{Explicit chiral symmetry breaking}

Explicit chiral symmetry breaking
due to the current quark masses is introduced through
\begin{eqnarray}
\hat{\chi} = 2 B \, \xi_L {\mathcal M} \xi_R^\dag
\ ,
\end{eqnarray}
where
${\mathcal M}$ is the quark mass matrix and $B$ is a constant
with dimension $1$.
The transformation property under the chiral symmetry is
\begin{eqnarray}
\hat{\chi} \ \to \ 
  h_L \, \hat{\chi}\, h_R
\ ,
\end{eqnarray}
where
$h_{L,R} \in [\mbox{SU}(N_f)_{L,R}]_{\rm local}$.
Symmetry breaking terms relevant to the meson masses are found 
as~\footnote{
In general, there are six independent terms including
the two of $\hat{\alpha}_{\parallel,\perp}$ and $\hat{\alpha}_M$
in one trace.  Here we use two terms which contribute to
the masses of vector and axialvector mesons.
Furthermore, we neglect the correction to the kinetic 
term of the gauge fields.
}
\begin{eqnarray}
&&
{\cal L}_{\rm \chi SB}
= \frac{h_V}{g^2}\, \mbox{tr} \left[
  \left(
    \hat{\alpha}_{\parallel}^\mu \hat{\alpha}_{\parallel \mu}
    + \hat{\alpha}_{\perp}^\mu \hat{\alpha}_{\perp \mu}
  \right)
  \left( \hat{\chi} \xi_M^\dag + \xi_M \hat{\chi}^\dag \right)
\right]
\nonumber\\
&&\quad
{}+ \frac{h_A - h_V}{g^2} \, \mbox{tr} \left[
  \hat{\alpha}_{M}^\mu \hat{\alpha}_{M \mu}
  \left( \hat{\chi} \xi_M^\dag + \xi_M \hat{\chi}^\dag \right)
\right]\,,
\label{lag sb}
\end{eqnarray}
with coefficients $h_V$ and $h_A$.
The additional piece~(\ref{lag sb}) in the Lagrangian gives
the meson masses and the pion decay constant as
\begin{eqnarray}
&&
M_\rho^2 = ag^2F^2 + h_V m_\pi^2\,,
\nonumber\\
&&
M_{a_1}^2 = \left( a + c \right)g^2F^2 + h_A m_\pi^2\,,
\nonumber\\
&&
F_\pi^2 = \left( cF^2 + \frac{h_A - h_V}{g^2}m_\pi^2 \right)
\frac{M_\rho^2}{M_{a_1}^2}\,,
\label{mr ma fp}
\end{eqnarray}
with non-zero pion mass $m_\pi$ and
to leading order in the symmetry breaking quark masses.
Flavor symmetry leads to the following relations in terms of 
light non-strange $(s=0)$ and strange meson masses
\begin{eqnarray}
&&
M_\rho^2 = ag^2F^2 + h_V m_\pi^2\,,
\nonumber\\
&&
M_{K^\ast}^2 = ag^2F^2 + h_V m_K^2\,.
\end{eqnarray}
One finds
\begin{equation}
h_V = \frac{M_{K^\ast}^2-M_\rho^2}{m_K^2-m_\pi^2}\,.
\end{equation}
The isospin $\frac{1}{2}$ states with $J^{PC} = 1^{+\pm}$ are
mixed. The $K_{1A}(1^{++})$ and $K_{1B}(1^{+-})$
are nearly equal mixtures of the $K_1(1270)$
and $K_1(1400)$ (with a $45^\circ$ mixing angle)~\cite{pdg}. 
Thus, the $h_A$ is expressed as
\begin{equation}
h_A = \frac{M_{K_{1A}}^2-M_{a_1}^2}{m_K^2-m_\pi^2}\,.
\end{equation}

In the present model, the coupling of $a_1$ to $\rho$-$\pi$ is 
determined by
\begin{equation}
g_{a_1\rho\pi} = - g^2 F_\pi \,,
\label{a1 rho pi coupling}
\end{equation}
where $F_\pi$ is given in Eq.~(\ref{mr ma fp}).
For expressing the $\rho$-photon mixing strength $g_\rho$ and 
$\rho$-$\pi$-$\pi$ coupling $g_{\rho\pi\pi}$
we introduce the higher derivative terms~\cite{hy:pr,hls:dl}. 
The resultant expressions are given by
\begin{eqnarray}
&&
g_\rho(s) = g \left( aF^2 + \frac{h_V}{g^2}m_\pi^2 - z_\rho s\right)\,,
\nonumber\\
&&
g_{\rho\pi\pi}(s) = \frac{g}{2}
\left( 1 + \frac{M_\rho^2}{M_{a_1}^2} - z_{\rho\pi\pi}
\frac{s}{F_\pi^2} \right)\,,
\end{eqnarray}
with the squared four-momentum $s = p^2$ and dimensionless
constants $z_{\rho}$ and $z_{\rho\pi\pi}$.
The parameters are fixed by comparison with experimental values
listed in Table~\ref{table}.
\begin{table*}
\begin{center}
\begin{tabular*}{17cm}{@{\extracolsep{\fill}}ccccccccc}
\hline
$F_\pi$ [GeV] &
$m_\pi$ [GeV] &
$m_K$ [GeV] &
$M_\rho$ [GeV] &
$M_{K^\ast}$ [GeV] &
$M_{a_1}$ [GeV] &
$M_{K_{1A}}$ [GeV] &
$g_\rho$ [GeV$^2$] &
$g_{\rho\pi\pi}$
\\
$0.0924$ &
$0.140$ &
$0.494$ &
$0.775$ &
$0.892$ &
$1.26$ &
$1.34$ &
$0.119$ &
$6.00$
\\
\hline
$aF^2$ [GeV$^2$] &
$cF^2$ [GeV$^2$] &
$g$ &
$h_V$ &
$h_A$ &
$z_\rho \times 10^3$ &
$z_{\rho\pi\pi} \times 10^3$
\\
$0.0133$ &
$0.0226$ &
$6.61$ &
$0.869$ &
$0.927$ &
$-7.09$ &
$-6.21$
\\
\hline
\end{tabular*}
\end{center}
\caption{
(Upper line)
Input quantities taken from PDG~\cite{pdg}. The values of
$g_\rho$ and $g_{\rho\pi\pi}$ are estimated from the decay widths
$\Gamma(\rho \to e^+e^-)$ and $\Gamma(\rho \to \pi\pi)$.
(Lower line)
Resulting model parameters.
}
\label{table}
\end{table*} 

\section{Chiral symmetry restoration}

The critical temperature $T_c$ for the restoration of chiral 
symmetry in its Wigner-Weyl realization
is defined as the temperature at which
the vector and axialvector current correlators coincide and 
their spectra become degenerate. Expanding the correlators (\ref{pole})
in the meson rest frame, one finds
\begin{equation}
G_A - G_V \propto M_\rho^2 (M_{a_1}^2 - M_\rho^2) 
= M_\rho^2 \delta M^2\,.
\end{equation}
Then chiral symmetry restoration implies
either $\delta M = 0$ or $M_\rho = 0$ (or both) at $T = T_c$:
Either the $\rho$-$a_1$ mass difference $\delta M$ or
the $\rho$ meson mass is identified
as a measure
of spontaneous chiral symmetry breaking and acts as an order
parameter of the chiral phase transition.

\subsection{Option A : dropping $a_1$ and non-dropping $\rho$ masses}

The GHLS theory describes the chiral symmetry restoration
with massless $\rho$ and $a_1$ mesons in the chiral limit~\cite{ghls}
(see also next subsection).
The classification of possible restoration patterns relies on
the renormalization group equations (RGEs). The theory does not 
have explicit scalar $\bar{q}q$ modes which will be important 
in the vicinity of the critical temperature. The scalar bosons 
may modify 
the RGEs and the massless mesons, protected by the fixed point of
the RGEs, might not necessarily be uniquely associated with the 
chiral symmetry restoration. 
This option suggests a symmetry restoration scenario in which
non-vanishing $\rho$ and $a_1$ masses become degenerate at
$T = T_c$.

For the case of non-dropping $\rho$ mass,
we will examine
$\delta M$ changing with temperature intrinsically such that 
$G_A - G_V = 0$ at the chiral transition.
To achieve $G_A = G_V$ with $\delta M = 0$ at the critical
temperature, we adopt the following ansatz of the 
temperature dependence of
the {\it bare} axialvector meson mass:
\begin{equation}
M_{a_1}^2 = M_\rho^2 + \delta M^2(T)\,,
\quad
\delta M^2(T) = c(T)g^2F^2\,,
\label{deltaM}
\end{equation}
with
\begin{eqnarray}
&&
c(T)
= c\,\Theta(T_f - T)
\nonumber\\
&&\qquad 
{}+ c\,\Theta(T - T_f)
\frac{T_c^2 - T^2}{T_c^2 - T_f^2}\,,
\nonumber\\
&&
g(T) = g\,,
\label{nodr ct}
\end{eqnarray}
where we schematically introduce the ``flash temperature'' 
$T_f$~\cite{BLR} which controls how the mesons experience 
partial restoration of chiral symmetry. 
The temperature dependence of $c(T)$ as well as the critical 
temperature $T_c$ are in principle determined by QCD,
e.g. through the matching to the QCD current correlators
of finite temperature.
We adopt here a simplified parameterization of the
$T$-dependence
\footnote{
 The pion decay constant near the critical temperature $T_c$
 behaves as $f_\pi^2 \sim T_c^2 - T^2$ in the chiral limit~\cite{hs:vm}.
 The parameterization in Eq.~(\ref{nodr ct}) describes this
 scaling.
 Here the $F_\pi$ denotes the tree-level parameter 
 given in Eq.~(\ref{tree}),
 while the $f_\pi$ indicates the physical quantity including
 hadronic corrections which are generated from loop
 diagrams at finite temperature.
}
in which
the values of $T_c$ and $T_f$ are taken in a reasonable range
as indicated, for example, by the onset of the chiral crossover
transition observed in lattice QCD~\cite{LGT}.
We take $T_c = 200$ MeV and $T_f = 0.7\,T_c$ for our numerical 
calculations.

For finite $m_\pi$ the temperature dependence of the $a_1$ meson
mass is given by
\begin{eqnarray}
M_{a_1}^2(T) = \left( a + c(T) \right) g^2 F^2
{}+ h_A m_\pi^2\,,
\label{nodrmpi}
\end{eqnarray}
where $m_\pi$ is assumed to be independent of temperature.

\subsection{Option B : dropping $a_1$ and $\rho$ masses}

The phase structure of the GHLS theory in vacuum was studied in 
detail based on the RG flows at one loop~\cite{ghls} assuming
that the scalar mesons are heavier than any other mesons 
and are integrated out near the critical point.
Here we give a brief summary of the chiral symmetry restoration
with massless $\rho$ and $a_1$ mesons in the GHLS.
In this case
chiral symmetry restoration can be realized only if the gauge 
coupling vanishes at the critical point,
\begin{equation}
 g \to 0\,,
\label{gtc}
\end{equation}
when one requires the first and second Weinberg sum rules to be 
satisfied. This option leads to the $\rho$ and $a_1$ mesons 
being massless:
\begin{equation}
M_\rho \to 0\,, 
\quad
M_{a_1} \to 0\,.
\end{equation}
The vanishing masses are not renormalized at the critical point
since $g=0$ is the only fixed point of its RGE.
This is a field theoretical description of
the dropping masses following Brown-Rho scaling~\cite{br}.
Possible patterns of the symmetry restoration are classified
by the mass ratio $M_\rho/M_{a_1}$ which flows into one of the 
following fixed points~\footnote{
 Besides (I) and (II), the fixed point 
 $M_\rho^2/M_{a_1}^2 \to 1/3$ also leads to a possible
 restoration pattern~\cite{ghls}. This is an ultraviolet fixed 
 point in any direction, so that it is not stable as to (I) and (II). 
 Thus, we will consider only type (I) and (II) in this paper.
}:
\begin{eqnarray}
&
\mbox{(I)}
&
\mbox{$\rho$-$a_1$ chiral partners} \ : \ 
  M_\rho^2/M_{a_1}^2 \ \rightarrow\ 1 \,,
\nonumber\\
&
\mbox{(II)}
&
\mbox{$\rho$-$\pi$ chiral partners} \ : \ 
  M_\rho^2/M_{a_1}^2 \ \rightarrow\ 0 \,.
\end{eqnarray}
These cases correspond to the Lagrangian parameters as
\begin{eqnarray}
&
\mbox{(I)}
&
  a \ \neq 0\,,
\quad
  c \ \rightarrow\ 0 \,,
\nonumber\\
&
\mbox{(II)}
&
  a \ \rightarrow\ 0 \,,
\quad
  c \ \neq 0\,.
\end{eqnarray}

The dropping $\rho$ and $a_1$ masses are described by the 
$T$-dependent gauge coupling $g(T)$ parameterized as~\cite{hls:dl}
\begin{equation}
g(T) = g\,\Theta(T_f - T) + g\,\Theta(T - T_f)
\sqrt{\frac{T_c^2 - T^2}{T_c^2 - T_f^2}}\,.
\label{gtdr}
\end{equation}
Two possible cases of chiral symmetry restoration are thus
distinguished by adopting the following parameterization with 
Eq.~(\ref{gtdr}):
\begin{eqnarray}
\mbox{(I)}\,:
&&
a(T) = a\,,
\nonumber\\
&&
c(T)
= c\,\Theta(T_f - T)
\nonumber\\
&&\qquad 
{}+ c\,\Theta(T - T_f)
\frac{T_c^2 - T^2}{T_c^2 - T_f^2}\,,
\label{gl}
\\
\mbox{(II)}\,:
&&
a(T) 
= a\,\Theta(T_f - T) 
\nonumber\\
&&\qquad
{}+ a\,\Theta(T - T_f)
\frac{T_c^2 - T^2}{T_c^2 - T_f^2}\,,
\nonumber\\
&&
c(T) = c\,.
\label{vm}
\end{eqnarray}

In the presence of explicit chiral symmetry breaking,
the $\rho$ and $a_1$ meson masses have the following temperature
dependence:
\begin{eqnarray}
\mbox{(I)}\,:
&& 
M_\rho^2(T) = a g^2(T)F^2 + h_V m_\pi^2\,,
\nonumber\\
&&
M_{a_1}^2(T) = (a + c(T)) g^2(T) F^2 + h_A m_\pi^2\,,
\label{glmpi}
\nonumber\\
\\
\mbox{(II)}\,:
&&
M_\rho^2(T) = a(T)g^2(T)F^2 + h_V m_\pi^2\,,
\nonumber\\
&&
M_{a_1}^2(T) = (a(T) + c) g^2(T) F^2 + h_A m_\pi^2\,,
\label{vmmpi}
\nonumber\\
\end{eqnarray}
with the scaling behaviors given in Eqs.~(\ref{gtdr}),
(\ref{gl}) and (\ref{vm}).

\section{Vector spectral function}

The vector current correlator Eq.~(\ref{CCdef}) in GHLS
is expressed in terms of two-point functions of the vector
gauge field $V^\mu$ and the external field ${\cal V}^\mu$
as~\cite{hs:vvd}
\begin{equation}
G_V = 
\frac{\Pi_V^S \left( \Pi_V^{LT} + 2\Pi_{V\parallel}^{LT}\right)}
{\Pi_V^S - \Pi_V^{LT}} + \Pi_\parallel^{LT}\,,
\label{VCC}
\end{equation}
where $\Pi_V, \Pi_{V\parallel}$ and $\Pi_\parallel$ are
$V$-$V$, $V$-${\cal V}$ and ${\cal V}$-${\cal V}$ 
correlation functions given explicitly in Appendix~\ref{app:loop},
with the following generic tensor decomposition:
\begin{equation}
\Pi^{\mu\nu} = g^{\mu\nu}\Pi^S 
{}+ \left( \frac{q^\mu q^\nu}{q^2} - g^{\mu\nu}\right) 
\Pi^{LT}\,.
\end{equation}
The vector spectral function is defined as the imaginary part
of the vector correlator in Eq.~(\ref{VCC}).

\subsection{Option A : dropping $a_1$ and non-dropping $\rho$ masses}

\begin{figure*}
\begin{center}
$T = 0.6\,T_c$ (below $T_f$)
\\
\includegraphics[width=8cm]{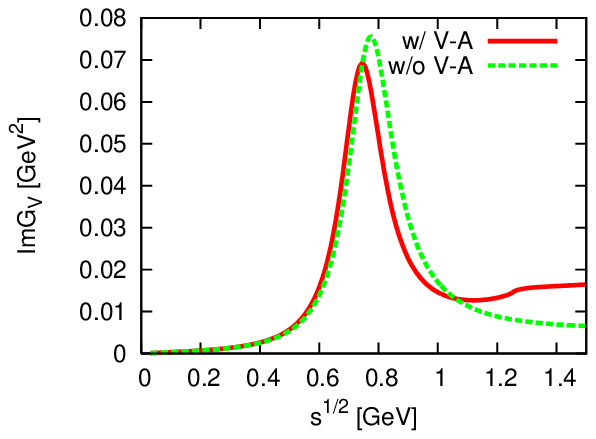}
\includegraphics[width=8cm]{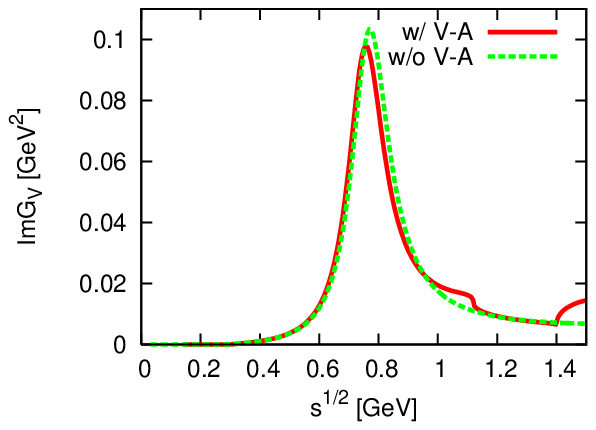}
\\
$T = 0.8\,T_c$ (above $T_f$)
\\
\includegraphics[width=8cm]{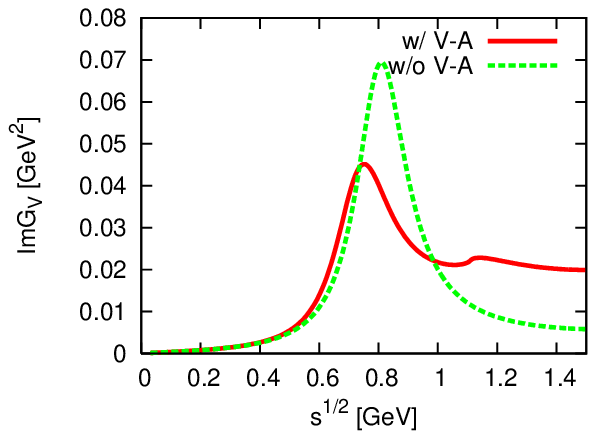}
\includegraphics[width=8cm]{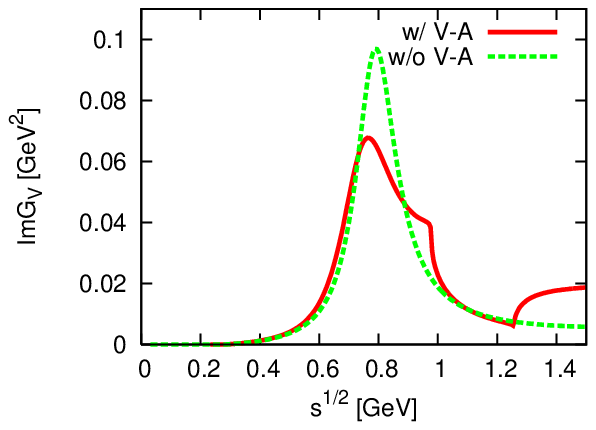}
\caption{
The vector spectral function for option A
at temperature $T/T_c = 0.6$ (upper) 
and at $T/T_c = 0.8$ (lower) with the critical temperature 
$T_c = 200$ MeV, calculated in the $\rho$-meson rest frame.
The left side figures are calculated for $m_\pi = 0$
and the right side for $m_\pi = 140$ MeV.
The solid curve is obtained in the full calculation. The dashed 
line is calculated eliminating the axialvector meson and hence
V-A mixing from the theory.
}
\label{vamix}
\end{center}
\end{figure*}

We first show, in the case of non-dropping $\rho$ mass,
the spectral function
in the chiral limit calculated in the GHLS theory 
in Fig.~\ref{vamix} (left).
Two cases are compared; one includes the V-A mixing and the other
does not. The spectral function has a peak at $M_\rho$ and a 
broad bump around $M_{a_1}$ due to the mixing. The height of the
spectrum at $M_\rho$ is enhanced and a contribution above 
$\sim 1$ GeV is gone when one omits the $a_1$ in the calculation.
One observes that a discrepancy between the two curves becomes
larger above $T_f$ where partial restoration of chiral symmetry
sets in.
For finite $m_\pi$ the energy of the virtual $\rho$ meson for
two processes, $\rho + \pi \to a_1$ and 
$\rho \to a_1 + \pi$, are splitted into $\sqrt{s} = m_{A_1} - m_\pi$
and $\sqrt{s} = m_{A_1} + m_\pi$.
This results in
the threshold effects seen as a shoulder at $\sqrt{s}=M_{a_1}- m_\pi$
and a bump above $\sqrt{s}=M_{a_1} + m_\pi$ 
in Fig.~\ref{vamix} (right).
Note that the enhancement of the spectrum for $m_\pi \neq 0$
is due to the change of the phase space factor $(s - 4m_\pi^2)^{3/2}$.

\begin{figure*}
\begin{center}
\includegraphics[width=8cm]{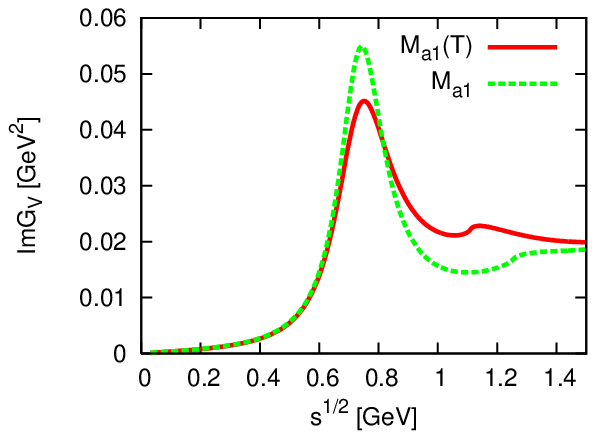}
\includegraphics[width=8cm]{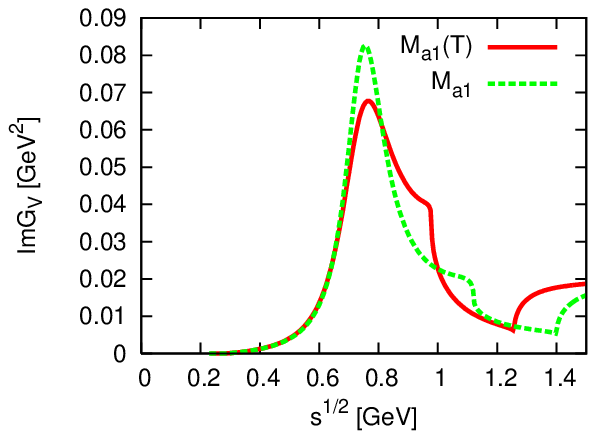}
\caption{
The vector spectral function at temperature
$T/T_c = 0.8$ for $m_\pi = 0$ (left) and
for $m_\pi = 140$ MeV (right).
The solid lines are obtained for Option A
where the $a_1$ mass has a temperature
dependence given in Eq.~(\ref{nodr ct}).
The dashed lines are calculated for a constant
$a_1$ mass.
}
\label{modification}
\end{center}
\end{figure*}

In Fig.~\ref{modification} we compare the vector spectrum
for Option A, where the $a_1$ bare mass changes with
temperature, with that for a constant bare mass.
Fig.~\ref{modification} (left) shows that
the upper bump due to the presence of $a_1$ appears 
at lower $\sqrt{s}$ than $M_{a_1}(T=0)=1.26$ GeV
since partial restoration of chiral symmetry sets in
which makes the $a_1$ mass decreasing.
In case of constant $a_1$ mass, 
this bump stays at the same point as vacuum $M_{a_1}$
at any temperatures.
The threshold effects for finite $m_\pi$ systematically
go down for the $T$-dependent $a_1$ mass, and show
no shift for the constant $a_1$ mass 
in Fig.~\ref{modification} (right).
The enhancement around $\sqrt{s} \lesssim 1$\,GeV will be
a signal of the partial chiral restoration.

\begin{figure*}
\begin{center}
\includegraphics[width=8cm]{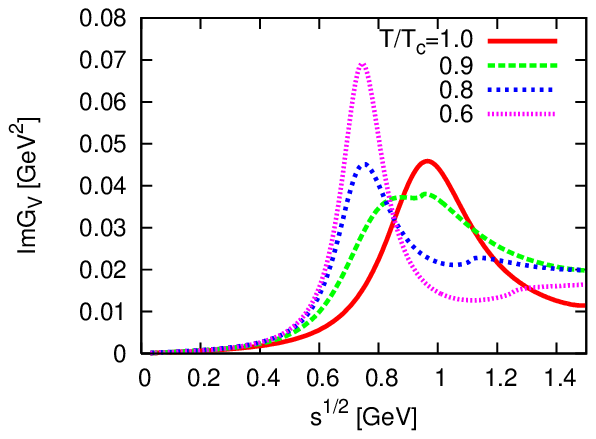}
\includegraphics[width=8cm]{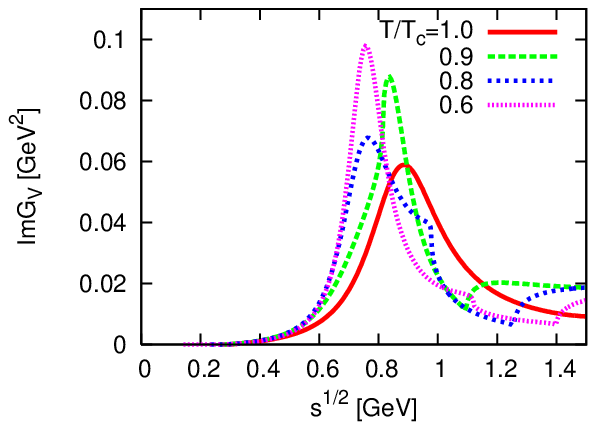}
\caption{
The vector spectral function (option A)
for $m_\pi = 0$ (left) and
for $m_\pi = 140$ MeV (right) at several temperatures 
$T/T_c = 0.6$-$1.0$.
}
\label{tdep}
\end{center}
\end{figure*}

Fig.~\ref{tdep} (left) shows the temperature dependence of the
vector spectral function in the chiral limit.
One observes a systematic downward shift of the enhancement
around the $a_1$ mass with temperature, 
while the peak position corresponding to the $\rho$ pole mass
moves upward due to the hadronic temperature corrections.
At $T/T_c = 0.9$ two bumps begin to overlap:
the lower one corresponds to the $\rho$ pole, and the upper one
to the $a_1$-$\pi$ contribution.
Finally at $T=T_c$, $M_{a_1}$ becomes degenerate 
with $M_\rho$ around $\sqrt{s} \simeq1\,$GeV
and the two bumps are on top of each other. 
Note that the V-A mixing eventually vanishes there.
This feature is a direct consequence of vanishing coupling of $a_1$ to 
$\rho$-$\pi$, as easily seen from Eq.~(\ref{a1 rho pi coupling}).
It is unchanged even if an explicit scalar 
field is present~\cite{hsw}.
Figure~\ref{tdep} (right) shows the temperature dependence of
the vector spectrum for finite $m_\pi$. Below $T_c$ one observes
the previously mentioned threshold effects moving downward with 
increasing temperature. It is remarkable
that at $T_c$ the spectrum shows almost no traces of 
$a_1$-$\rho$-$\pi$ threshold effects:
Eq.~(\ref{nodrmpi}) together with the fact that $h_V \simeq h_A$
shows that the $\rho$ to $a_1$ mass ratio becomes almost $1$ at 
$T=T_c$:
\begin{equation}
\frac{M_\rho^2}{M_{a_1}^2}
\stackrel{T \to T_c}{\to}
\frac{ag^2F^2 + h_V m_\pi^2}{ag^2F^2 + h_A m_\pi^2}
\simeq 1\,,
\label{ratio}
\end{equation}
and the pion decay constant is very tiny there,
$F_\pi^2 \sim (h_A-h_V) m_\pi^2 / g^2$.
Consequently, Eq.~(\ref{a1 rho pi coupling}) implies that
$g_{a_1 \rho\pi} \sim \sqrt{ h_A - h_V }\, m_\pi \sim 0.06\, m_\pi$.
This
indicates that at $T_c$ {\it the $a_1$ meson mass nearly equals
the $\rho$ meson mass and the $a_1$-$\rho$-$\pi$ coupling
almost vanishes even in the presence of explicit
chiral symmetry breaking.}

\subsection{Option B : dropping $a_1$ and $\rho$ masses}

\begin{figure*}
\begin{center}
\includegraphics[width=8cm]{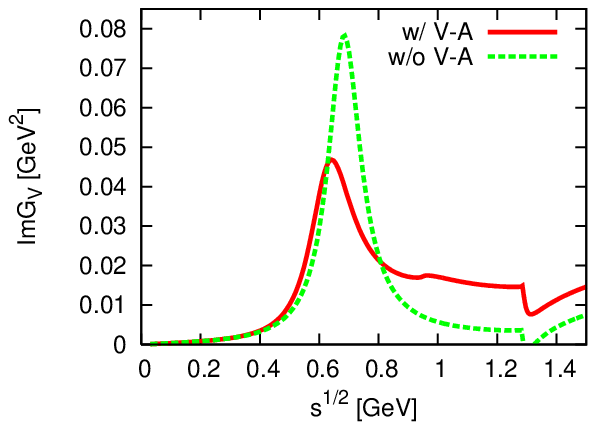}
\includegraphics[width=8cm]{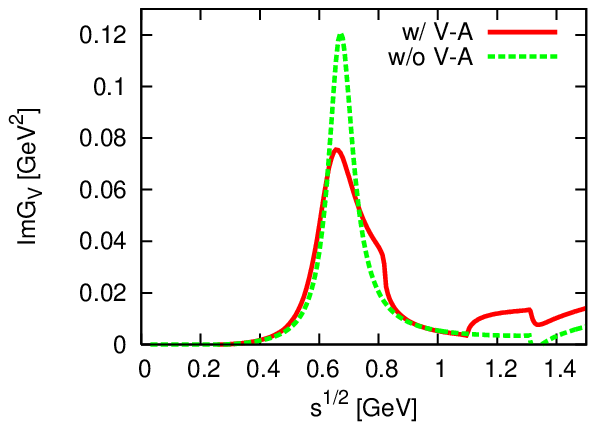}
\caption{
The vector spectral function (option B)
for $m_\pi = 0$ (left)
and for $m_\pi = 140$ MeV (right) in type (I)
at temperature $T/T_c = 0.8$ with the critical temperature 
$T_c = 200$ MeV, calculated in the $\rho$-meson rest frame. 
}
\label{glva}
\end{center}
\end{figure*}

In case of dropping $\rho$ and $a_1$ masses,
the spectral function is enhanced compared to that without dropping
mass since the $\rho$ decay width is reduced~\cite{hls:dl}.
Fig.~\ref{glva} shows the vector spectrum using the type (I) 
parameterization at $T = 0.8\,T_c$. 
The feature that the $a_1$ meson suppresses the vector spectral function
through the V-A mixing remains unchanged~\cite{hsw:qm}.
Compared with Fig.~\ref{vamix} (lower-left),
a bump through the V-A mixing and the $\rho$ peak are shifted downward 
since both the $\rho$ and $a_1$ masses drop.
The self-energy has a cusp at the threshold $2\,M_\rho$ and
this appears as a dip at $\sqrt{s} \sim 1.3$ GeV.
The influence of finite $m_\pi$ turns out to be in threshold
effects as before.

\begin{figure*}
\begin{center}
\includegraphics[width=8cm]{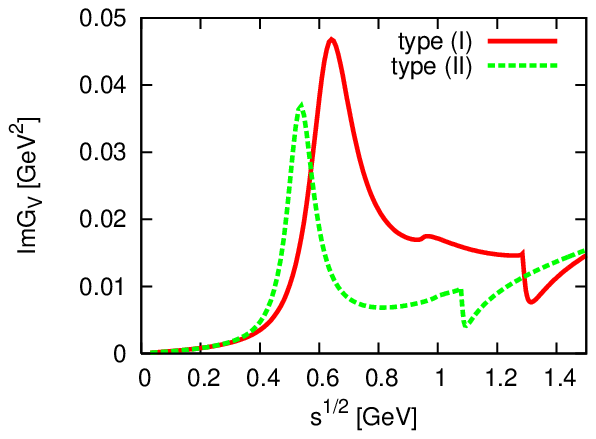}
\includegraphics[width=8cm]{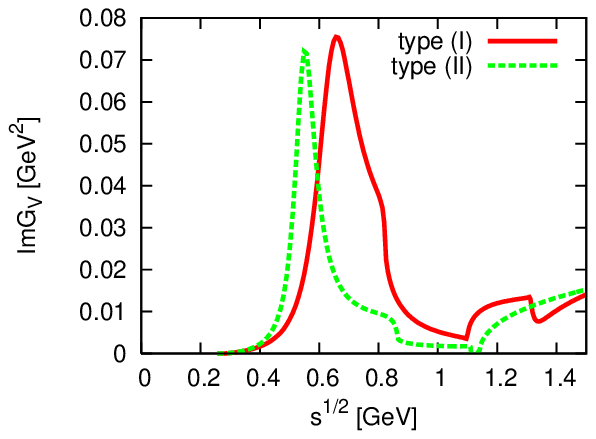}
\caption{
The vector spectral function (option B)
for $m_\pi = 0$ (left) and for $m_\pi = 140$ MeV (right)
at temperature $T/T_c = 0.8$ with the critical temperature 
$T_c = 200$ MeV, calculated in the $\rho$-meson rest frame. 
The solid curve is for type (I). The dashed line is for type (II). 
}
\label{glvm}
\end{center}
\end{figure*}

In Fig.~\ref{glvm} we compare type (I) with (II) at
$T = 0.8\,T_c$.
In type (II) the $\rho$ meson mass drops faster than the $a_1$ mass
which is clearly seen in the figure. The $\rho$ coupling to the
vector current $g_\rho$ decreases faster than that for type (I)
and this makes the spectral function somewhat suppressed compared
with that for type (I).

For finite pion mass, one finds from Eqs.~(\ref{glmpi}) and
(\ref{vmmpi}) the mass ratio near $T_c$
\begin{equation}
\frac{M_\rho^2}{M_{a_1}^2} \stackrel{T \to T_c}{\to}
\frac{h_V}{h_A} \simeq 1\,,
\end{equation}
for {\it both} type (I) and (II). This leads to the nearly 
vanishing V-A mixing as seen for the non-dropping $\rho$ mass,
option A (see Eq.~(\ref{ratio})).

It should be noted that the vector meson becomes the chiral partner
of the pion and vector meson dominance is strongly violated
when the chiral symmetry is restored in the VM (type (II))~\cite{vm}. 
This induces a significant reduction of the vector spectral 
function~\cite{hls:dl,bhhrs}. On the other hand, the pion form
factor is still vector-meson dominated at $T_c$ if the dropping 
$\rho$ and $a_1$ join in the same chiral multiplet 
(type (I))~\cite{ghls}.

\section{Conclusions}

We have performed a detailed study of V-A mixing in the current
correlation functions and its evolution with temperature,
guided by three possible scenarios of chiral symmetry restoration:
dropping $\rho$ and $a_1$ masses with type (I) and (II), 
and alternatively dropping $a_1$ mass becoming degenerate with
a non-zero $\rho$ meson mass at critical temperature.
In the chiral limit the axialvector meson contributes significantly 
to the vector spectral function; the presence of the $a_1$ 
reduces the vector spectrum around $M_\rho$
and enhances it around $M_{a_1}$.
For physical pion mass $m_\pi$, the $a_1$ contribution 
above $\sqrt{s} \sim M_{a_1}$ still survives although the bump is 
somewhat reduced.
A major change with both dropping $\rho$ and $a_1$ masses is
a systematic downward shift of the vector spectrum.
We observe a different evolution of the spectrum depending on
type (I) or (II) before reaching the critical temperature.
The $a_1$-$\rho$-$\pi$ coupling vanishes at the critical temperature
$T_c$ and thus the V-A mixing also vanishes.
A remarkable observation is that even for physical $m_\pi$ 
the $\rho$ and $a_1$ meson masses are well degenerate at $T_c$. 
The vanishing V-A mixing at $T_c$ stays almost intact.

One interesting application of this thermal spectral function
is to study dilepton production in relativistic heavy-ion collisions.
The change of the V-A mixing in the presence of matter and 
its influence on dilepton production has been evaluated based 
on a virial 
expansion for $T < m_\pi$ and $\rho < 3\rho_0$ (with normal 
nuclear matter density $\rho_0$)~\cite{chreduction}. However, 
important modifications of the $a_1$-meson properties near 
critical temperature have not been treated so far
in dilepton processes in the context of chiral symmetry restoration.
Of course, in order to deal with dileptons realistically 
one needs to account for
other collective excitations and many-body interactions 
as well as the time evolution of the created fireball~\cite{dileptons}. 
Such effects can screen signals of chiral restoration~\cite{bhhrs} 
and make an interpretation of broad in-medium
spectral functions in terms of a changing chiral order parameter
quite difficult~\cite{qcdsr}.
The situation at RHIC and/or LHC might be very different from
that at SPS. At SPS energies many-body effects come from
the presence of baryons. These effects are expected to be
much reduced in very
hot matter with relatively low baryon density. 
The present study may then be of some relevance for the high
temperature, low baryon density scenarios encountered
at RHIC and LHC.

One caveat in the present treatment is about the lack of genuine
$\bar{q}q$ scalar which becomes the chiral partner of the pion
in the Ginzburg-Landau picture of chiral symmetry restoration.
The scalar modes are expected to be important near the chiral
critical temperature and may modify the current correlators.
This can be quantified by introducing explicit
scalar modes in a GHLS invariant way. 
Work concerning the finite temperature 
evolution of both vector and axialvector spectral functions 
in this generalized framework is in progress and
will be reported elsewhere~\cite{hsw}.

\subsection*{Acknowledgments}

The work of C.~S. and W.~W. has been supported in part by BMBF and 
by the DFG cluster of excellence ``Origin and Structure of the 
Universe''.
The work of M.H. has been supported in part by
the JSPS Grant-in-Aid for Scientific Research (c) 20540262
and Global COE Program 
``Quest for Fundamental Principles in the Universe''
of Nagoya University provided by Japan Society for the
Promotion of Science (G07).

\begin{widetext}

\appendix

\setcounter{section}{0}
\renewcommand{\thesection}{\Alph{section}}
\setcounter{equation}{0}
\renewcommand{\theequation}{\Alph{section}.\arabic{equation}}

\section{Two-point functions at one-loop}
\label{app:loop}

A systematic derivative expansion based on the GHLS was adopted
in Ref.~\cite{ghls} where one finds details of its construction 
and quantization procedure. In the following, we list the expressions
for three relevant two-point functions.

We define the Feynman integrals by
\begin{eqnarray}
&&
A_0(M)
= T\sum_{n=-\infty}^\infty
\int \frac{d^3 k}{(2\pi)^3}\frac{1}{M^2 - k^2}\,,
\nonumber\\
&&
B_0(p;M_1,M_2)
=  T\sum_{n=-\infty}^\infty
\int \frac{d^3 k}{(2\pi)^3}
  \frac{1}{[M_1^2-k^2][M_2^2-(k-p)^2]}\,, 
\nonumber\\
&&
B^{\mu\nu}(p;M_1,M_2)
=  T\sum_{n=-\infty}^\infty
\int \frac{d^3 k}{(2\pi)^3}
  \frac{(2k-p)^\mu (2k-p)^\nu }{[M_1^2-k^2][M_2^2-(k-p)^2]}\,,
\end{eqnarray}
where the 0th component of the loop momentum is taken as
$k^0 = i2n\pi T$ and that of the external momentum
$p^0 = i2n^\prime\pi T$ [$n,n^\prime$: integer]
in the standard Matsubara formalism.

The two-point function of the vector gauge field $V_\mu$ is
given by
\begin{eqnarray}
\Pi_V^{\mu\nu}
&=&
\int d^4x\,e^{ipx}
\langle T\,V^\mu(x)V^\nu(0) \rangle
\nonumber\\
&=& N_f \zeta g^{\mu\nu}A_0(m_\pi)
{}+ 2N_f g^{\mu\nu} A_0(M_\rho)
{}+ N_f \left( \zeta^2 - 2\zeta + 3 \right) A_0(M_{a_1})
\nonumber\\
&&
{}+ \frac{N_f}{8}(1+\zeta)^2 B^{\mu\nu}(p;m_\pi,m_\pi)
\nonumber\\
&&
{}- N_f\left[M_\rho^2 g^{\mu\nu}-4(p^2 g^{\mu\nu}-p^\mu p^\nu)\right]
B_0(p;M_\rho,M_\rho)
{}+ \frac{9N_f}{8}B^{\mu\nu}(p;M_\rho,M_\rho)
\nonumber\\
&&
{}- N_f\left[M_\rho^2\zeta g^{\mu\nu}
{}- 4(p^2 g^{\mu\nu}-p^\mu p^\nu)\right] B_0(p;M_{a_1},M_{a_1})
{}+ \frac{N_f}{8}(\zeta^2 - 4\zeta + 12) B^{\mu\nu}(p;M_{a_1},M_{a_1})
\nonumber\\
&&
{}- N_f M_\rho^2 (1-\zeta) g^{\mu\nu} B_0(p;M_{a_1},m_\pi)
{}+ \frac{N_f}{4}\zeta(1-\zeta) B^{\mu\nu}(p;M_{a_1},m_\pi)\,,
\end{eqnarray}
where we introduce a temperature-dependent parameter $\zeta$ as
\begin{equation}
\zeta(T) = 
\begin{cases}
\frac{M_\rho^2}{M_{a_1}^2(T)}
& \text{for dropping $a_1$ and non-dropping $\rho$ (option A)}
\\
\frac{M_\rho^2(T)}{M_{a_1}^2(T)}
& \text{for dropping $a_1$ and $\rho$: type (I) or (II) (option B)}
\end{cases}\,.
\end{equation}
The relevant one-loop diagrams to the V-A mixing are shown in
Fig.~\ref{diagram}.
\begin{figure}
\begin{center}
\includegraphics[width=8cm]{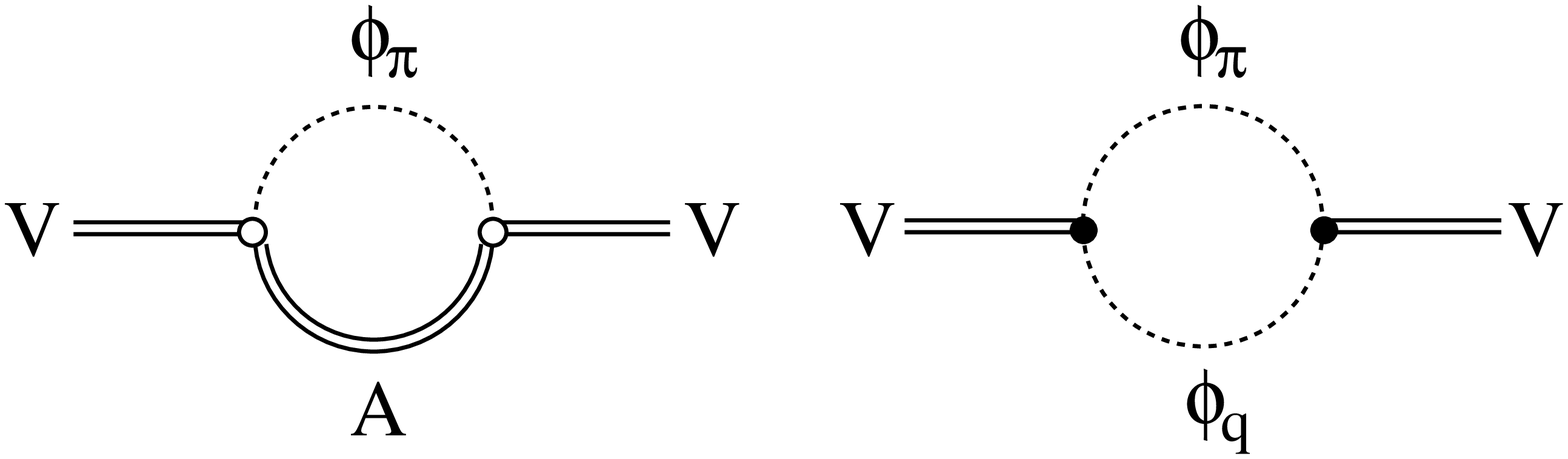}
\caption{
Diagrams contributing to the V-A mixing at one loop. 
The circle $(\circ)$ denotes the momentum-independent 
vertex and the dot $(\bullet)$ denotes the momentum-dependent vertex.
Vector and axialvector fields are denoted by $V$ and $A$
and pion by $\phi_\pi$.
The $A$ is the transverse components of $a_1$ meson, while
the $\phi_q$ the longitudinal one.
}
\label{diagram}
\end{center}
\end{figure}
The left diagram is proportional to $B_0(p;M_{a_1},m_\pi)$
and the right to $B^{\mu\nu}(p;M_{a_1},m_\pi)$.
One easily finds that the V-A mixing
generated from those diagrams vanishes at the critical temperature
{\it independently of the pattern of chiral restoration}, i.e.,
type (I): $\zeta = 1$, type (II): $\zeta = 0$ 
for $M_{a_1} = M_\rho = 0$, or $\zeta = 1$
for $M_{a_1} = M_\rho \neq 0$ at $T_c$.

The two-point function of $V^\mu$ and the external vector field
${\cal V}^\nu$, like a photon, is found as
\begin{eqnarray}
\Pi_{V\parallel}^{\mu\nu}
&=&
\int d^4x\,e^{ipx}
\langle T\,V^\mu(x){\cal V}^\nu(0) \rangle
\nonumber\\
&=& \frac{N_f}{2}(1-\zeta) g^{\mu\nu}A_0(m_\pi)
{}+ \frac{N_f}{2}g^{\mu\nu} A_0(M_\rho)
{}+ \frac{N_f}{2}\zeta g^{\mu\nu} A_0(M_{a_1})
\nonumber\\
&&
{}+ \frac{N_f}{8}(1-\zeta^2) B^{\mu\nu}(p;m_\pi,m_\pi)
\nonumber\\
&&
{}+ N_f M_\rho^2 g^{\mu\nu}B_0(p;M_\rho,M_\rho)
{}+ \frac{N_f}{8}B^{\mu\nu}(p;M_\rho,M_\rho)
\nonumber\\
&&
{}+ N_f M_\rho^2 \zeta g^{\mu\nu}B_0(p;M_{a_1},M_{a_1})
{}+ \frac{N_f}{8}\zeta(2-\zeta) B^{\mu\nu}(p;M_{a_1},M_{a_1})
\nonumber\\
&&
{}+ N_f M_\rho^2 (1-\zeta) g^{\mu\nu}B_0(p;M_{a_1},m_\pi)
{}- \frac{N_f}{4}\zeta(1-\zeta) B^{\mu\nu}(p;M_{a_1},m_\pi)\,.
\end{eqnarray}

The two-point function of ${\cal V}^\mu$ is
\begin{eqnarray}
\Pi_\parallel^{\mu\nu}
&=&
\int d^4x\,e^{ipx}
\langle T\,{\cal V}^\mu(x){\cal V}^\nu(0) \rangle
\nonumber\\
&=& 
\frac{N_f}{8}(1-\zeta)^2 B^{\mu\nu}(p;m_\pi,m_\pi)
\nonumber\\
&&
{}- \frac{N_f}{8}M_\rho^2 g^{\mu\nu}B_0(p;M_\rho,M_\rho)
{}+ \frac{N_f}{8}B^{\mu\nu}(p;M_\rho,M_\rho)
\nonumber\\
&&
{}- N_f M_\rho^2 \zeta B_0(p;M_{a_1},M_{a_1})
{}+ \frac{N_f}{8}\zeta^2 B^{\mu\nu}(p;M_{a_1},M_{a_1})
\nonumber\\
&&
{}- N_f M_\rho^2 (1-\zeta) g^{\mu\nu} B_0(p;M_{a_1},m_\pi)
{}+ \frac{N_f}{4}\zeta(1-\zeta) B^{\mu\nu}(p;M_{a_1},m_\pi)\,.
\end{eqnarray}

\end{widetext}


\end{document}